\documentclass[twocolumn,showpacs,aps,prb,superscriptaddress,floatfix]{revtex4}
\usepackage{graphicx}
\usepackage{dcolumn}
\usepackage{bm}

\begin{document}

\preprint{}


\affiliation{Department of Physics and Astronomy, McMaster University, Hamilton, Ontario, L8S 4M1, Canada}
\affiliation{Canadian Institute for Advanced Research, 180 Dundas St. W., Toronto, Ontario, M5G 1Z8, Canada}

\author{J.P. Castellan}
\affiliation{Department of Physics and Astronomy, McMaster University, Hamilton, Ontario, L8S 4M1, Canada}
\author{B.D. Gaulin}
\affiliation{Department of Physics and Astronomy, McMaster University, Hamilton, Ontario, L8S 4M1, Canada}
\affiliation{Canadian Institute for Advanced Research, 180 Dundas St. W., Toronto, Ontario, M5G 1Z8, Canada}
\author{H.A. Dabkowska}
\affiliation{Department of Physics and Astronomy, McMaster University, Hamilton, Ontario, L8S 4M1, Canada}
\author{A. Nabialek}
\affiliation{Institute of Physics, Polish Academy of Sciences, Al. Lotnikow 32/46, 02-668, Warsaw, Poland}

\author{G. Gu}
\affiliation{Department of Physics, Brookhaven National Laboratory,  Upton, N.Y., 11973-5000, U.S.A.}

\author{Z. Islam}
\affiliation{Advanced Photon Source, Argonne National Laboratory, Argonne, IL, 60439, U.S.A.}
\author{X. Liu}
\affiliation{Department of Physics, University of California, San Diego,CA, 92093, U.S.A.}
\author{S.K. Sinha}
\affiliation{Department of Physics, University of California, San Diego,CA, 92093, U.S.A.}

\title{Two and Three Dimensional Incommensurate Modulation in Optimally-Doped Bi$_2$Sr$_2$CaCu$_2$O$_{8+\delta}$}

\begin{abstract}
X-ray scattering measurements on optimally-doped single crystal samples of the high temperature superconductor Bi$_2$Sr$_2$CaCu$_2$O$_{8+\delta}$ reveal the presence of three distinct incommensurate charge modulations, each involving a roughly fivefold increase in the unit cell dimension along the {\bf b}-direction.  The strongest scattering comes from the well known (H, K$\pm$ 0.21, L) modulation and its harmonics.  However, we also observe broad diffraction which peak up at the L values complementary to those which characterize the known modulated structure.  These diffraction features correspond to correlation lengths of roughly a unit cell dimension, $\xi_c$$\sim$20 $\AA$ in the {\bf c} direction, and of $\xi_b$$\sim$ 185 $\AA$ parallel to the incommensurate wavevector.  We interpret these features as arising from three dimensional incommensurate domains and the interfaces between them, respectively. In addition we investigate the recently discovered incommensuate modulations which peak up at (1/2, K$\pm$ 0.21, L) and related wavevectors.
Here we explicitly study the L-dependence of this scattering and see that these charge modulations are two dimensional in nature with weak correlations on the scale of a bilayer thickness, and that they correspond to short range, isotropic correlation lengths within the basal plane.  We relate these new incommensurate modulations to the electronic nanostructure observed in Bi$_2$Sr$_2$CaCu$_2$O$_{8+\delta}$ using STM topography.
\end{abstract}

\newcommand{\bscco}{Bi$_2$Sr$_2$CaCu$_2$O$_8$$_+$$_\delta$ }

\maketitle 

\section{Introduction}
High temperature superconductivity is observed and has been extensively studied across several families of layered copper oxides\cite{Reviews}. While the 
structure of families of these materials differ in detail, they are all comprised of stackings of CuO$_2$ rectangular layers, 
with so-called charge resevoir layers in between.  The CuO$_2$ layers can occur in isolation, or appear in bi- or trilayer assemblies.  
One of the most extensively studied of these materials is \bscco, which is known to display an optimal superconducting T$_C$ of $\sim$ 93 K \cite{Reviews}.  
As shown in Fig. 1, it is a bilayer superconductor with orthorhombic symmetry and lattice parameters of a$\sim$b$\sim$ 5.4 $\AA$, and c=30.8 $\AA$ \cite{Fischer}.  As such two CuO$_2$ bilayers appear within the unit cell separated by $\sim$ 15.4 $\AA$ along {\bf c}.

\bscco has been the high temperature superconducting sample of choice for studies using surface-sensitive techiques such as photoemmission\cite{Damascelli} and scanning tunneling microscopy (STM)\cite{Davis_materials}.  This is due to the relative ease of producing clean, cleaved surfaces of \bscco at low temperature, a consequence of its extreme quasi-two dimensionality, even compared with other families of layered high temperature superconductors, such as YBa$_2$Cu$_3$O$_{7-\delta}$ and La$_{2-x}$Sr$_x$CuO$_4$.  These techniques and the information they provide on electronic structure have been very important to the general field of high temperature superconductivity, and \bscco has been crucial to these studies.  Unfortunately, the two dimensional nature of the material has also resulted in relatively thin single crystals being grown (with thicknesses in the {\bf c} direction of as much as, but typically smaller than, a few millimeters).  This has precluded, or made very difficult, the study of the \bscco family by experimental techniques which require a large volume of material, such as single crystal neutron scattering\cite{Keimer}.

\begin{figure}[t]
\centering
\includegraphics[width=7cm]{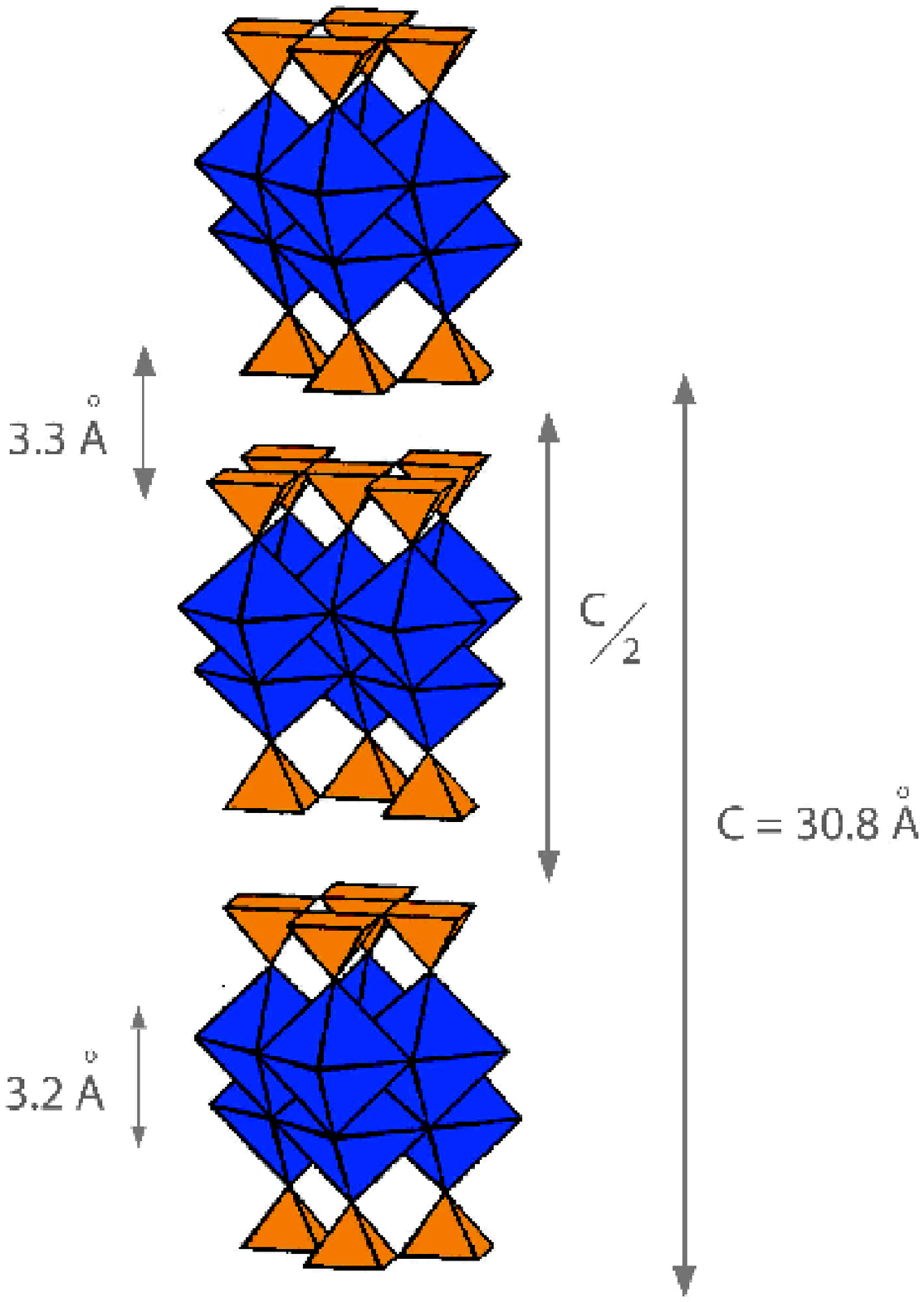}
\caption{A schematic diagram depicting one and a half unit cells of \bscco is shown.  The light (orange) pyramids represent CuO$_2$ plus apical oxygen complexes, 
while the dark (blue) octahedra represent BiO$_2$ plus apical oxygen atom complexes.  The Cu ions reside at the center of the base of the light pyramids, while the Bi 
ions reside at the center of the octahedra.  The CuO$_2$ and BiO$_2$ bilayer thicknesses are indicated as 3.3 $\AA$ and 3.2 $\AA$, respectively, while the 
bilayer-bilayer separation of c/2 $\sim$ 15.4 is also shown.} 
\label{unitcell} 
\end{figure}

The structural detail for which \bscco is perhaps best appreciated is the incommensurate modulation it possesses within its basal plane.  This incommensurate modulation has been observed by different techniques including, x-ray\cite{xray}, neutron\cite{neutron} and electron diffraction\cite{electron}; scanning tunneling microscopy\cite{STM}, and electron microscopy\cite{elect_microscopy}. This complex crystal structure has been described in terms of two interpenerating periodic crystals, each of which modulate the other\cite{Walker}.

The modulation is one-dimensional, running along the {\bf b}-direction in the basal plane, and it is the modulation (as opposed to a difference between a and b lattice parameters) that is responsible for the lower-than-tetragonal symmetry of the structure.  The incommensurate modulation is characterized by multiple harmonics, and the first order harmonic corresponds to a periodicity of roughly 5 unit cells along {\bf b}.  Interestingly, it appears that \bscco crystals typically grow untwinned, likely a consequence of strains which would develop between orthogonal modulations.  This is in contrast
to other families of high temperature superconductors which can be produced in an untwinned state, but only with considerable effort in preparation.

\section{Materials and Experimental Methods}

The \bscco single crystal samples were grown by the travelling solvent floating zone techniques using a Crystal Systems Inc. image furnace. The growth was performed from premelted polycrystalline rods in 2 atmospheres of flowing O$_2$.  The rods were counter-rotated at 30 rpm and the growth speed was 0.2 mm/hour.  The resulting single crystals were carefully extracted from a large boule, and had approximate dimensions of 10 mm $\times$ 5 mm $\times$ 0.1 mm, where the thin dimension was parallel to the {\bf c} axis.  Although thin, the single crystals were of excellent quality with mosaic spreads less than 0.07 degrees, and a superconducting transition temperature, determined by SQUID magnetomery, of T$_C$$\sim$ 92 K with a width of $\sim$ 2 K.

Two sets of x-ray scattering measurements were performed on these samples. Measurements were made using a Cu-rotating anode source and a triple axis 
spectrometer at McMaster University.  These measurements employed Cu-K$\alpha$ radiation and utilized a pyrolitic graphite monochromator.  The sample was mounted in a closed cycle refrigerator and placed in a four-circle goiniometer.  Four sets of slits were employed to define the incident and scattered beams.  

A high resolution diffraction setup at the Advanced Photon Source synchrotron radiation facility was used to study a subset of the modulations with greater precision.  
These measurements were performed on the 4-ID-D beamline at an x-ray energy of 16 keV.  This beamline is equipped with a double-crystal Si(111) monochromator.  A Pd-coated bent mirror was used to focus the beam to a spot size of 180 $\times$ 220 $\mu$m at the sample.  A flat
Pd mirror was used to further suppress higher order harmonics.  The thin single crystal sample was placed in a closed cycle refridgerator.  A Ge(111) analyser was used to reduce background and to improve resolution.   

\section{Experimental Results}

\subsection{Three Dimensional Modulated Structure}

\begin{figure}[h]
\centering
\includegraphics[width=7cm]{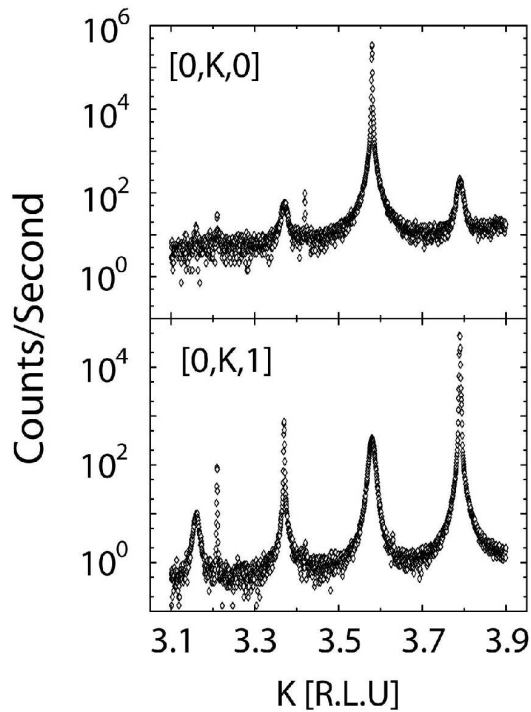}
\caption{Synchrotron X-ray scattering scans on a logarithmic intensity scale along [0,K,0] (top panel) and [0,K,1] (lower panel) in \bscco.  A total of four harmonics of the modulated structure with the form (0, 4-n$\times\tau$, L) are observed at both L=0 and L=1, and they clearly alternate between resolution-limited and broad along both K and L (see Fig. \ref{3dlscan}).  The broad component to the scattering is well described by a Lorentzian-squared lineshape (see Fig. \ref{fittolorsq}).} 
\label{3dkscan} 
\end{figure}

Synchrotron x-ray scattering scans of the form (0,K,0) and (0,K,1) are shown in Fig. \ref{3dkscan}.  This data is plotted on a logarithmic intensity scale, 
and was taken at T=10 K, although no significant temperature dependence was observed to this scattering to temperatures as high as 300 K.  
A clear pattern of diffraction peaks is observable at incommensurate wavevectors of the form (0, 4-n$\times\tau$, L) for $\tau$=0.21, corresponding to a roughly 5 unit cell modulation, and at all harmonics (n=1,2,3,4) which we studied.  Two distinct lineshapes are clearly present.  One is very sharp, resolution limited, and these peaks alternate in harmonic for a particular L.  This is to say that the even harmonic, n=2 (K=3.58), is clearly sharp in K for L=0, while it is the odd harmonics, n=1 (K=3.79) and n=3 (3.37), which are sharp for L=1. 

\begin{figure}[h]
\centering
\includegraphics[width=7cm]{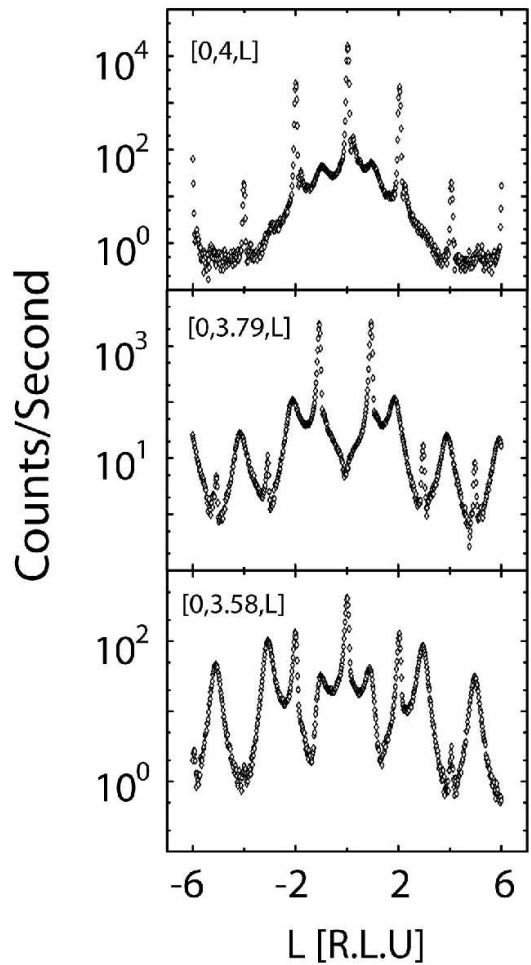}
\caption{Rotating anode x-ray scattering measurements on a logarithmic intensity scale show the L-dependence of the principal Bragg peaks, such as [0,4,L] (top panel); the first order harmonic of the modulated structure [0, 3.79, L] (middle panel); and the second order harmonic of the primary modulated structure [0, 3.58, L] (bottom panel).  The harmonics alternate with L between resolution-limited Bragg scattering and broad scattering which is well described by a Lorentzian-squared lineshape (see Fig. \ref{fittolorsq}).} 
\label{3dlscan}
\end{figure}

Figure \ref{3dlscan} shows x-ray scattering data, plotted on a logarithmic intensity axis, taken with the rotating anode source.  
This data examines the systematic L dependence of the scattering shown in Fig. \ref{3dkscan}. Once again this data is taken at 10 K and 
scans of the form (0, 4, L), (0, 3.79, L), and (0, 3.58, L) are shown in the top, middle and bottom panels of Fig. \ref{3dlscan}, respectively.  Consistent with the data shown in Fig. \ref{3dkscan}, two sets of incommensurate diffraction peaks are seen in the bottom two panels, which alternate with L, between a sharp, resolution-limited lineshape, and a broad diffraction feature.

\begin{figure}
\centering
\includegraphics[width=7cm]{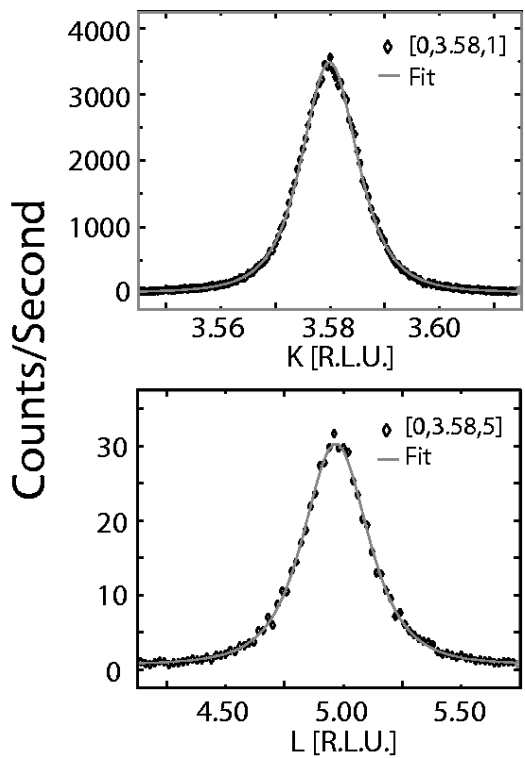}
\caption{X-ray scattering scans along K (top) and L (bottom) taken from Figs. \ref{3dkscan} and \ref{3dlscan} and plotted on a linear intensity scale.
Both plots show the second order harmonics of the primary modulation [0, 3.58, 1] and [0, 3.58, 5].  These positions correspond to wavevectors with broad 
(as opposed to resolution-limited) lineshapes.  Also shown is the fit to each lineshape using a Lorentzian-squared form, Eq. 1, which is clearly excellent.  As described in the text, this lineshape is typically used to describe the three dimensional random field domain state in disordered magnets.}  
\label{fittolorsq} 
\end{figure}

Clearly, at alternate harmonics one observes a much broader lineshape.  This is considered quantitatively in Fig. \ref{fittolorsq}, where broad diffraction features at (0, 3.58, 1) and (0, 3.58, 5) are plotted as a function of K and L on a linear intensity scale.  This data was fit to a Lorentzian-squared form along both {\bf b$^*$} and {\bf c$^*$}:

\begin{equation}
S({\bf Q})  =  {A\over{{[1+{({\bf Q}-{\bf Q_o})^2\over{\kappa}^2}]}^2}}
\end{equation}

yielding correlation lengths of $\xi_b={1\over{\kappa_b}}\sim$ 185 $\AA$ and $\xi_c$=20 $\AA$.

Together these data illustrate two co-existing incommensurate structures one of which displays resolution-limited long range order in three dimensions, while the other displays three dimensional, albeit highly anisotropic, short range order.  We interpret the anisotropic, short-range order diffraction peaks as arising from stacking faults in the incommensurate modulation.  

For simplicity, we discuss this three dimensional scattering as if a single harmonic characterized the incommensurate structure.  Domains of well  
ordered modulated regions give rise to the resolution limited peaks at say (0, 3.79, L) with L=odd.  The L=odd requirement implies a phase shift of of 
$\pi$, 3$\times\pi$, 5$\times\pi$, between either CuO$_2$ bilayers (or equivalently BiO$_2$ bilayers) which are separated from each other by half a unit cell 
along {\bf c} (see Fig. 1).  That is, the modulated structure within a particular bilayer is $\pi$ out of phase with that immmediately above and below it.  

In contrast the L-dependence of the broad diffraction peaks, such as (0, 3.79, L) with L=even, implies that this incommensurate modulation is in 
phase with that in the bilayer immediately above and below it.  Also, as the correlation length along {\bf c} is roughly a unit cell dimension, we 
have a very natural interpretation that these diffraction features arise from the interface between two bulk domains, but which are out of phase with 
respect to each other.  While the domain interfaces, attributed to the broad diffraction peaks, are anisotropic with $\xi_c$ $\sim$ 20 $\AA$ and 
$\xi_a$ $\sim$ 185 $\AA$, we classify all of the (0, 4 $\pm$ n$\times$$\tau$, L) superlattice diffraction peaks as being three dimensional in nature, as all of the superlattice peaks are localized around a particular point in reciprocal space, as opposed to being organized into rods or sheets of scattering, as is known to be relevant to two and one dimensional structures, respectively.   

\subsection{Two Dimensional Modulated Structure}

\begin{figure}[h]
\centering
\includegraphics[width=7cm]{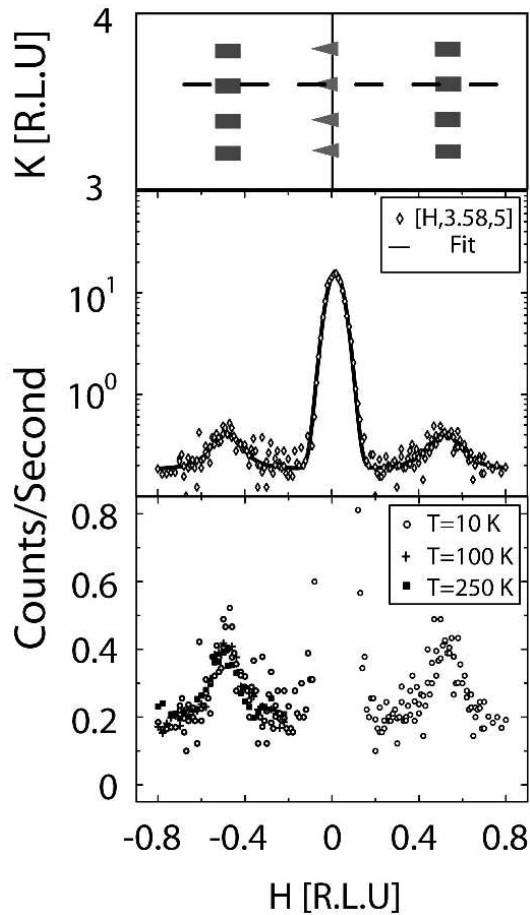}
\caption{X-ray scattering scans of the form [H, 3.58, 5] which show the new modulated structure corresponding to a doubling of the unit cell along {\bf a} is shown.  
The dashed line in the upper panel shows where the scans reside in reciprocal space, passing through the second order harmonic of the 
primary modulation.  The middle and lower panels show these scans on logarithmic (middle) and linear (bottom) intensity scales, with Lorentzian lineshapes (Eq. 2)
 fit to the data.  The bottom panel also demonstrates the lack of temperature dependence to this scattering below room temperature.}
\label{2dhscan}
\end{figure}

X-ray scattering measurements using the rotating anode source were also extended to look for additional charge modulation.  Additional modulation, distinct from that 
at (0, 4-n$\times\tau$, L) and discussed above, involving a doubling of the periodicity along {\bf a} have recently been reported
in \bscco by Megtert and co-workers\cite{Megtert}, and also by ourselves\cite{ourselves}.  Scans performed along H, going through the second order 
harmonic of the three dimensional modulation, (0, 4-2$\times$0.21, 5), are shown in the bottom two panels of Fig. \ref{2dhscan}.  The (0, 3.58, 5) second order 
harmonic of the three dimensional modulation is easily observable at the $\sim$ 10 counts/second level.  However, we also observe peaks at 
($\pm$1/2, 3.58, 5) at the 0.4 counts/second level on a $\sim$ 0.2 counts/second background. While weak, this H=1/2 modulation is clearly resolved in the middle panel of Fig. \ref{2dhscan}.  The bottom panel of Fig. \ref{2dhscan} shows three data sets at T=10 K, 100 K, and 250 K, at the (-1/2, 3.58, 5) wavevector and  demonstrates that, like the H=0 modulations discussed above, the new H=1/2 modulations have little or no temperature dependence to temperatures as high as room temperature.  The peaks shown in Fig. \ref{2dhscan} were fit to a resolution-convoluted Lorentzian form:  
 
\begin{figure}[t]
\centering
\includegraphics[width=7cm]{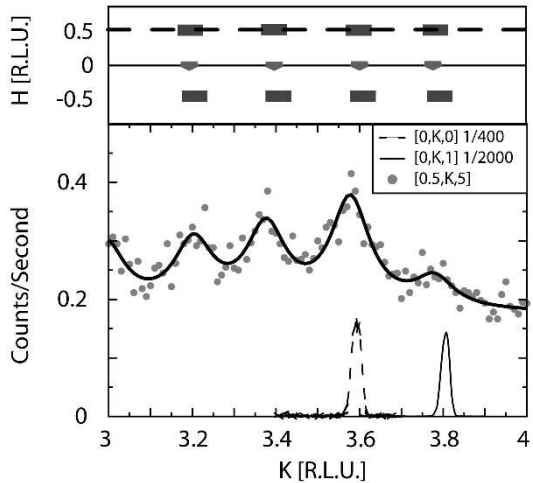}
\caption{X-ray scattering scans of the form [1/2, K, 5] which display the new modulated structure corresponding to a doubling of the unit cell along {\bf a} is shown.  
The upper panel shows where the scans reside in resiprocal space, passing through the first four harmonics of this new modulated structure.  
The data in the bottom panel is fit to Lorentzian lineshapes (Eq. 2) centered at each of the four harmonic positions along {\bf K}.  Also shown for comparison in the same panel are {\bf K} scans of the first and second harmonics of the primary modulated structure, whose linewidth defines the instrumental resolution.}
\label{2dkscan}
\end{figure}

\begin{equation}
S({\bf Q})  =  {A\over{{[1+{({\bf Q}-{\bf Q_o})^2\over{\kappa}^2}]}}}
\end{equation}

and the ($\pm$1/2, 3.58, 5) peaks were found to be characterized by a correlation length of $\xi_a$ $\sim$ 27 $\AA$ along {\bf a}.

Figure \ref{2dkscan} shows x-ray scattering scans of the form (1/2, K, 5) passing between K=3 and 4, and showing (1/2, 4-n$\times\tau$, 5) for $\tau$=0.21 and n=1,2,3, and 4.  The top panel of Fig. \ref{2dkscan} locates these scans in reciprocal space as the dashed lines.  Also shown in the bottom panel of Fig. \ref{2dkscan}, are scans of the form (0, K, 0) and (0, K, 1) taken under precisely the same experimental conditions as the (1/2, K, 5) scan.  This data shows K scans through (0, 3.58, 0) and (0, 3.79, 1) which are both resolution-limited Bragg peaks (see Fig. \ref{3dkscan}).  We therefore conclude that the peaks shown in the (1/2, K, 5) scans of Fig. \ref{2dkscan} are not resolution limited.  We can provide an excellent fit to this data, shown as the solid line in Fig. \ref{2dkscan}, with a resolution-convoluted Lorentzian lineshape, Eq. 2.  We then extract a relatively short correlation length along {\bf b} for this H=1/2 modulation of $\xi_b$ $\sim$ 27 $\AA$, very similar to the correlation length $\xi_a$, extracted from the (H, 3.58, 5) scans of Fig. \ref{2dhscan}.  

\begin{figure}[t]
\centering
\includegraphics[width=7cm]{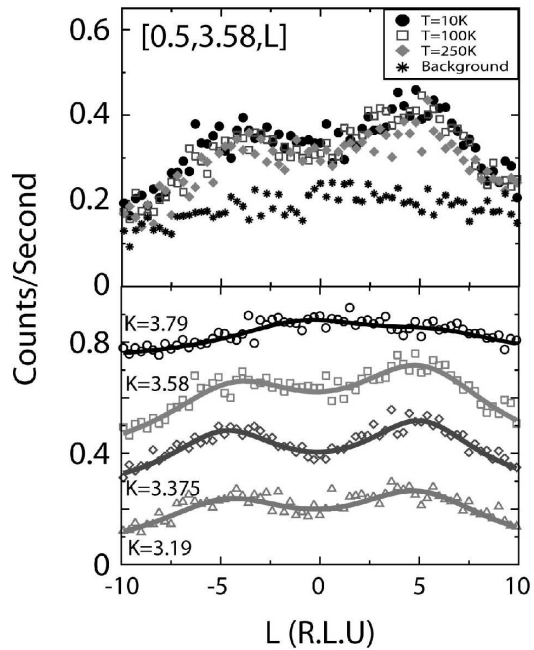}
\caption{X-ray scattering scans of the form [1/2, K, L] which observe the new modulated structure corresponding to a doubling of the unit cell along {\bf a} are shown.  The top panel shows L-scans of the second order (K=3.58) harmonic of this scattering, at different temperatures below 250 K, as 
well as a background scan of the form [.3, 3.45, L].  The lower panel shows the L dependence of each of the first four harmonics of this structure at 10 K.  Remarkably over the 10 Brillouin zones from L=10 to L=-10, a near complete rod of scattering is observed for the first order harmonic (K=3.79), while higher order harmonics peak up weakly at L=$\pm$5.}
\label{2dlscan}
\end{figure}
Figure \ref{2dlscan} shows L-scans of the form (1/2, 4-n$\times\tau$, L) for n=2 in the top panel, that is (1/2, 3.58, L), and for n=1, 2, 3, and 4 in the bottom panel. 
Also shown in the top panel is a background scan taken at (0.3, 3.45, L).  Remarkably these scans, from L=-10 to 10, show almost no L-dependence at all; 
that is rods of scattering characteristic of two dimensional structures.  The scattering is weakly peaked at L $\sim$ $\pm$ 5, but extends over all L values measured.  
These scans were also carried out as a function of temperature, and again, as seen in the top panel of Fig. \ref{2dlscan}, one sees little or no 
temperature dependence to temperatures as high as room temperature.  These results on the new (1/2, 4-n$\times\tau$, L) modulations extend those reported 
earlier\cite{Megtert, ourselves} by explicitly determining L dependencies.  We also see that our results give approximately isotropic 
correlation lengths within the {\bf a-b} plane a factor of two smaller than those reported by Megtert and collaborators\cite{Megtert}.  However, the Megtert et al. work does not discuss the form to which the scattering was fit to extract their correlation length, and hence a quantitative comparison is not possible at present.

The extreme extended nature of this scattering in L helps us understand why the scattering is so weak at any one position, and therefore why it is 
difficult to observe, and has escaped observation until recently.  As the scattering is broad in both H and K (with correlation lengths $\xi_{ab}$ 
$\sim$ 27 $\AA$ and rod-like in L, the integrated intensity of these H=1/2 incommensurate modulations is in fact quite appreciable.  However the scattering is sufficiently diffuse so a to make its peak intensity very weak, down by a factor of 10$^6$ compared with principal Bragg peaks of the three 
dimenensional structure, such as (0, 4, 0).   

\section{Discussion and Relation to STM Imaging}

Our results show three distinct sets of incommensurate modulation in optimally-doped \bscco, which co-exist at all temperatures below room
temperature.  Two of these are three dimensional in nature, with scattering centered on definite Bragg points in reciprocal space.  This 
scattering is peaked up at (H, 4-n$\times\tau$, L) and alternates as a function of harmonic at a particular L, and as a function
of L at a particular harmonic, between resolution-limited peaks and those characterized by finite and anisotropic correlation lengths in all three directions.

The broad, (H, 4-n$\times\tau$, L), incommensurate peaks are extremely well described by a Lorentzian-squared lineshape (Fig. \ref{fittolorsq}).  
This is an interesting result in itself, as a sum of Lorentzian plus Lorentzian-squared terms is typically used to model diffuse magnetic scattering in random 
field Ising model (RFIM) magnets, which display random local symmetry breaking\cite{Cowley}.  Three dimensional RFIM systems, such as Co$_x$Zn$_{1-x}$F$_2$\cite{CoZnF} 
and  Fe$_x$Zn$_{1-x}$F$_2$\cite{FeZnF} in the presence of an applied magnetic field, show magnetic diffraction features with negligible Lorentzian weight, and thus 
their diffuse scattering is also very well described by a Lorentzian-squared form.  Such a scattering form implies a pair-correlation function in three dimensions 
which falls off with distance as exp(-$\kappa$r), and is characteristic of a domain wall state\cite{Cowley}.  Our results for $\kappa _{ab}$ imply 
these domains are hundreds of angstroms across, with domain walls which are a single unit cell in extent in the {\bf c} direction.  This small spatial extent, 
$\xi_c$ $\sim$ 20 $\AA$ in the {\bf c} direction is consistent with local stacking defects.  As \bscco is more generally the m=2 
member of the Bi$_2$Sr$_2$Ca$_{m-1}$Cu$_m$O$_{4+2m+\delta}$ family, the nature of the stacking fault is likely a missing Ca-layer.  As such the domain wall itself is expected to  locally be the m=1 member of the family: Bi$_2$Sr$_2$CuO$_{6+\delta}$.  This is a relatively low T$_C$ superconductor (T$_C$$\sim$ 10 K) with a smaller c-axis, c=24.6 $\AA$\cite{Ginsberg} than \bscco.  Such a scenario has recently been proposed from high resolution 
electron microscopy measurements performed on the same crystals as those studied here\cite{Botton}.

The most remarkable result we report is the confirmation and characterization of the incommensurate modulation corresponding to a doubling of the periodicity along H 
and giving rise to rods of scattering of the form (1/2, 4-n$\times\tau$, L) at all L.  The scattering tends to be weakly peaked near L=$\pm$ 5.  	This indicates 
out-of-phase, or $\pi$, correlations {\it within} a bilayer.  As shown schematically in Fig. 1, the CuO$_2$ and BiO$_2$ bilayer thicknesses are 
almost identical, at $\sim$ 3.3 $\AA$.  Such correlations give rise to weak peaks at wavevectors of $\pi$/(bi-layer thickness) 
$\sim$ $\pi$/3.3 $\AA$ =0.95 $\AA ^{-1}$ $\sim$ L=5$\times$2$\pi$/c.  As can be seen in Fig. \ref{2dlscan}, this tendency for the two dimensional rods of 
scattering to peak up near L$\sim \pm$ 5 seems to be strongest for the second and third order harmonics of this incommensurate structure (1/2, 3.58, L) and 
(1/2, 3.375, L), and is not seen at all for the 1st order harmonic (1/2, 3.79, L) which appears to be truly rod-like in nature.

Several interesting questions arise from the observation of this two dimensional incommensurate structure.  First, from what does it arise? For example, is it a 
surface effect, or can it arise from an intrinsically two dimensional charge inhomogeniety which exists throughout the full volume of the crystal?

The scattering geometry employed in all of the measurements presented in this paper is {\it transmission} geometry, and hence it is very unlikely that the rod-like incommensurate scattering we observe in Figs. \ref{2dhscan}, \ref{2dkscan}, and \ref{2dlscan} is due to surface structure of any kind, even though it is two dimensional in nature.  We are then left with the intriguing conclusion that the scattering originates from the bulk of the crystal, but is nonetheless two dimensional in nature.


\begin{figure}[b]
\centering
\includegraphics[width=8.5cm]{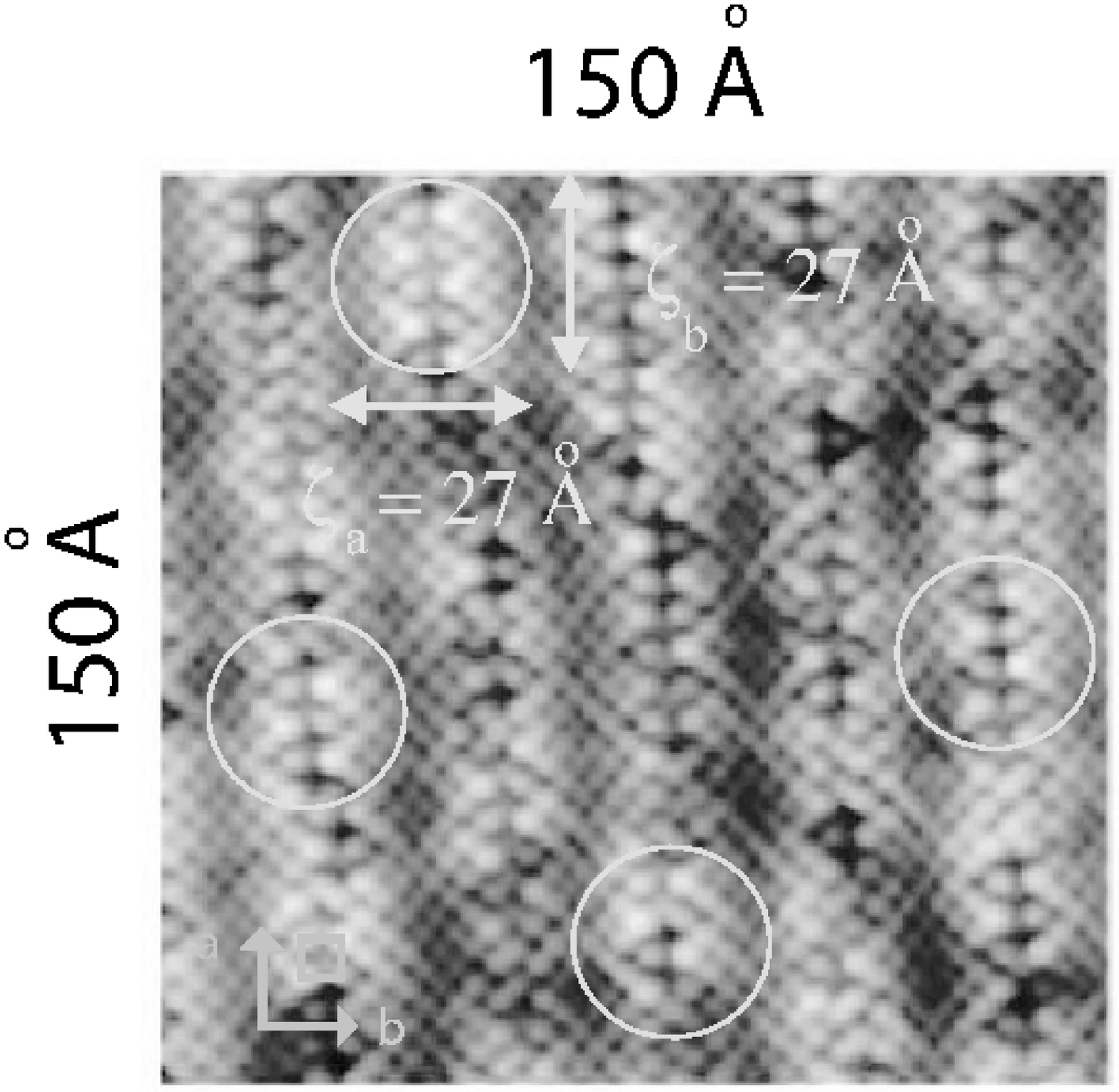}
\caption{The correlation areas of the new modulated structure corresponding to a doubling of the unit cell along {\bf a} extracted from our x-ray scattering analysis are superposed on STM topographs\cite{STM} of the surface of \bscco.  Our measurements indicate isotropic short range correlations with a correlation length of $\sim$ 27 $\AA$, and these correspond very well to the extent of the ``S" microstructure which appears to be imprinted on the primary modulated structure of \bscco, as observed in real space with the STM measurements.}
\label{stmpic}
\end{figure}
A second question which arises is, given that the \bscco family of superconductors have been studied for more than 15 years, has this type of structure been observed in previous experiments?  We believe the answer to this question is yes, and reconsider STM images of the surface structure of \bscco discussed several years ago by Pan and co-workers\cite{STM}.  Figure \ref{stmpic} shows an STM image of the [001] cleaved surface of near-optimally doped \bscco from Pan et al\cite{STM}.  The well known (0, 0.21, L) modulation is seen as the periodic lighter stripes which run along the vertical direction in the STM image and which have an approximate periodicity of 5 unit cells dimensions along {\bf b}.  However, superposed on the light stripes is an ``S-like" dark modulation, which has a periodicity of 2 unit cells along the stripe; that is along the {\bf a}-direction.  This additional modulation in the 
structure appears superposed on every light stripe within the field of view. As such one expects its manifestation in reciprocal space to appear 
at wavevectors of the form (1/2, 0.21, L), exactly where we observe the two dimensional rods of scattering.

Further, careful examination of the STM image shows that the S-modulation along {\bf a} is not a perfect pattern, but is characterized by phase slips both along {\bf a} and between S-modulations imposed on adjacent light stripes.  Consequently, such substructure would not give rise to resolution-limited Bragg peaks, but rather to diffuse peaks characterized by relatively short correlation lengths.  Again, this is fully consistent with the two dimensional rods of scattering which we have observed at (1/2, 4-n$\times\tau$, L), wherein the correlation length measured in the ab-plane was found to be approximately isotropic, $\xi_{ab}$ $\sim$ 27 $\AA$.  We have superposed correlation-areas within the ab plane using the correlation lengths extracted from our x-ray scattering measurements, onto the STM images\cite{STM} of the structure of the cleaved [001] plane of optimally-doped \bscco, and the agreement between the two is remarkably good. The S-modulation within the STM image was not commented on by Pan et al\cite{STM}.  However, as STM is a very surface sensitive probe, it provides little information as to the behaviour of this substructure into the bulk of the material.  It would therefore be possible to associate such structure in an STM experiment with a surface effect.  However, we reiterate that our x-ray scattering measurements, performed in transmission geometry, are very {\it insensitive} to the surface.  We conclude that the two dimensional modulated structure we observe in x-ray scattering is the same as the electronic inhomogneiety observed with STM, and that it originates from charge inhomogeniety within the bulk of the \bscco superconductor.

Finally, one can ask how general is the phenomenon of charge inhomogeniety in different families of high temperature superconductors?  Of course, electronic phase separation is implied within a variety of models for high temperature superconductivity, most prominently those involving ``stripes"\cite{stripes}.  However, is there related diffuse scattering evidence in other high temperature superconductors?

The answer again appears to be yes, with the recent reports of a four unit cell superstructure with correlation lengths of $\sim$ 20 $\AA$ within the ab plane and very short range correlations along {\bf c} in optimally doped YBa$_2$Cu$_3$O$_{7-\delta}$\cite{IslamPRL}, and related superstructure in underdoped YBa$_2$Cu$_3$O$_{7-\delta}$\cite{IslamPRB}. In these cases, the short range ordered superstructure is periodic along the shorter Cu-Cu bond, whereas the two unit cell superstructure we report here for \bscco is periodic along a diagonal of the CuO$_2$ basal plane cell, and thus at 45 degrees to either Cu-Cu bond.  This difference, however, may simply be a consequence of the electronic inhomogeniety following the appropriate template for 
\bscco or YBa$_2$Cu$_3$O$_{7-\delta}$.  That template would be set by the three dimensional modulation in \bscco and by the short range oxygen-vacancy ordered superstructures in the case of YBa$_2$Cu$_3$O$_{7-\delta}$\cite{IslamPRB}, which are present throughout the superconducting region of its phase diagram\cite{deFontaine}.  It may therefore be the case that such quasi-two dimensional, short range charge inhomogeneity is a general property of cuprate high temperature superconductors.
 
\section{Conclusions}

We report the observations of three sets of incommensurate modulation to the structure of the high temperature superconductor \bscco.  Two of these are the well known modulation characterized by wavevectors (0, 4-n$\times\tau$, L).  These Bragg features alternate in both harmonic, n, for the same L, and in L for the same harmonic, between resolution-limited Bragg peaks and diffuse scattering lineshapes which are very well described by a Lorentzian-squared profile.  We also confirm and extend the characterization of the recently discovered (1/2, 4-n$\times\tau$, L) modulations.  These charge correlations are quasi-two dimensional.  The rod-like character of this scattering along {\bf c$^*$} is weakly peaked at L=5, and this scattering is characterized by finite and relatively short correlation lengths of $\sim$ 27 $\AA$ within the ab plane.  We contend that the charge inhomogeniety underlying this diffuse scattering is that previously observed in STM measurements\cite{STM}.  These new x-ray scattering measurements, 
in transmission geometry, establish that this charge inhomogeniety is a bulk property of the system, and is not due to any reconstruction or related surface effect.  We have also thoroughly investigated any possible temperature dependence to all three of these structural modulations below room temperature and find no such effect.

\section{Acknowledgements}

We wish to acknowledge useful discussions with J.M. Tranquada. This work was supported by NSERC of Canada.  Use of the Advanced Photon Source is supported by the U.S. Department of Energy, Office of Basic Energy Sciences, under contract No. W-31-109-ENG-38.  This work was supported in part by the BES (DOE) Grant DE-FG02-03ER46084 (SKS).


\end{document}